\documentstyle[aasms4,flushrt]{article}

\begin{document}

\title{COMPARING THE EVOLUTION OF THE GALAXY DISK SIZES WITH CDM MODELS:
THE HUBBLE DEEP FIELD}
\author{E. Giallongo$^1$, N. Menci$^1$, F. Poli$^1$, S. D'Odorico$^2$, A. Fontana$^1$
}

\bigskip

\affil{$^1$ Osservatorio Astronomico di Roma, via di Frascati 33, I-00040
Monteporzio, Italy\\
\noindent
$^2$ European Southern Observatory, Karl Schwarzschild Strasse 2,
D-85748 Garching, Germany\\
}

\begin{abstract}

The intrinsic sizes of the field galaxies with $I\leq 26$ in the
Hubble and ESO-NTT Deep Fields are shown as a function of their
redshifts and absolute magnitudes using photometric redshifts derived
from the multicolor catalogs and are compared with the CDM
predictions. Extending to lower luminosities and to higher $z$ our
previous analysis performed on the NTT field alone, we find that the
distribution of the galaxy disk sizes at different cosmic epochs is
within the range predicted by typical CDM models.  However, the
observed size distribution of faint ($M_B>-19$) galaxies is skewed
with respect to the CDM predictions and an excess of small-size disks
($R_d<2$ kpc) is already present at $z\sim 0.5$. The excess persists
up to $z\sim 3$ and involves brighter galaxies . Such an excess may be
reduced if luminosity-dependent effects, like starburst activity in
interacting galaxies, are included in the physical mechanisms
governing the star formation history in CDM models.

\end{abstract}


\section{INTRODUCTION}

The understanding of galaxy formation has recently undergone an
appreciable progress. Observationally, this is due to photometric
and spectroscopic information in deep galaxy fields.  A corresponding
theoretical progress has been achieved by the developement of
semi-analytical approaches including the gas cooling and star formation
processes into the well developed hierarchical clustering theories for
dark matter (DM) halos; this allows to relate the galaxy DM circular
velocities to observable, luminous properties. In this context, the
Tully-Fisher (TF) relation represents a typical test for the present
theoretical models, as it relates the total luminosity of a disk
galaxy to its halo circular velocity.

However, from an observational point of view, the measure of circular
velocity is limited to bright nearby spirals and a few very
bright galaxies at intermediate redshifts (e.g. Vogt et al. 1997).
Extending the TF relation to fainter/distant spirals is essential to
test the evolution in the $L$, $z$ plane of the $M/L$ ratio predicted
by CDM theories.

An alternative statistical approach to connect the luminous and
dynamical properties of galaxy disks is based on the size-luminosity
relation, once a specific model is assumed to connect size to circular
velocity. This has the advantage of exploring the dynamical evolution
over a wide range of luminosities and redshifts.

In a previous paper (Poli et al. 1999, Paper I) we applied this novel
approach to the ESO-NTT Deep Field (Arnouts et al. 1999) to derive
morphological information for the galaxies in the field down to $I=25$
after appropriate seeing deconvolution.  The derived intrinsic angular
sizes were then converted into physical sizes adopting photometric
redshifts for each galaxy in the catalog (Fontana et al. 1999a). The
distribution of sizes in the $L$, $z$ plane were compared with
predictions of CDM semi-analytic models of galaxy formation (e.g. Cole
et al. 1994, Baugh et al. 1998) complemented with the specific
size-velocity relation worked out by Mo, Mao, \& White (1998) for
rotationally supported disks. This analysis showed an excess of
small-size low luminosity galaxies at small-intermediate
redshifts. However, the sample magnitude limit did not allow an assessment
of this excess
at higher redshifts. Here we want
to extend the study to higher $z$ and lower $L$ using the available
data in the Hubble Deep field North where morphological information on
the faint galaxy sample is available in the literature (Abraham et
al. 1996; Odewahn et al. 1996; Driver et al. 1998; Marleau \& Simard
1999). This will allow to assess if the excess observed at
intermediate redshifts is an evolutionary effect or if it is present
also at higher $z$, indicating the presence of remarkable physical
processes not included in the standard CDM models.

\section{THE GALAXY CATALOG}

The morphological information on the galaxies in the Hubble Deep Field
north was obtained using the galaxy catalog by Marleau \& Simard
(1999) where the structural parameter values were derived from
the intensity profile fitting. Exponential profiles were assumed for
the disk component and a final characteristic disk radius in arcsec
was computed for all the galaxies up to $I\simeq 26$.

The HDF-N galaxy catalog has been joined with the NTT Deep Field
catalog used in Paper I limited to $I\simeq 25$.

To each galaxy in the joined catalog, a photometric redshift estimate
was assigned with the same best fitting procedure applied in
Paper I and in Giallongo et al. 1998. This was obtained through a
comparison of the observed colors with those predicted by spectral
synthesis models (Bruzual \& Charlot 1996) including UV absorption by
the intergalactic medium and dust reddening. The aperture magnitudes
used to estimate the galaxy colors and the I total magnitude for each
galaxy were extracted from the catalog by Fernandez-Soto et al. (1999). The
catalog of the photometric redshifts is presented in Fontana et
al. 1999a. The resulting typical redshift accuracy is $\Delta z \simeq
0.06$ up to $z\sim 1.5$ and $\Delta z\sim 0.3$ at larger redshifts.

As derived in other fields of similar depth, the bulk of the galaxies
is concentrated at intermediate redshifts $z\sim 0.5-1$ with a tail in the
distribution up to $z\sim 6$.

In order to verify the effects of the background noise on the measured
sizes of the faint galaxies in the HDF we have performed a set of
simulations specifically designed to reproduce the typical conditions
of real data. The same test has been performed in Paper I for the
galaxies in the NTT Deep Field.

The intensity profiles were reproduced as in the observed HDF images
assuming the same pixel sampling.  Assuming an intrinsic exponential
profile, a series of synthetic images were constructed using different
half light radii ranging from $r_{hl}=0.1$'' to $r_{hl}=0.9$'' with a
step of 0.1''. An average of 25 random objects were computed for each
radius, assuming a total magnitude of $I\sim 26$ which is the limiting
magnitude of the morphological galaxy sample. The background
subtracted image of a bright star was selected in the field to
reproduce the instrumental PSF. Its normalized profile was then
convolved with the synthetic images of the disk galaxies.  The
convolved two dimensional profiles were randomly inserted in regions
of the HDF far from very bright objects to reproduce the observed HDF
galaxies with the appropriate pixel size and noise levels. Finally,
the multigaussian deconvolution technique was applied to the synthetic
data as in Paper I.

We first notice that there is no selection bias against galaxies with
large size ($r_{hl}\sim 1$ arcsec) and low surface brightness down to
$I\simeq 26$ since all the synthetic objects were detected.  The
results are shown in Fig.~1, where the error bars represent the
dispersion around the mean due to noise in the background
subtraction. A good match between the intrinsic and measured half
light radii was obtained up to $r_{hl}\sim 0.7$ arcsec. For
larger values, a slight underestimate in the measured values appears
at the sample limiting magnitude.  In any case it can be seen that,
even for the faint galaxies with $I\sim 26$, the overall
correlation between intrinsic and measured half-light radii is
preserved in such a way that an intrinsically large, faint object,
e.g. with $r_{hl}\sim 0.7''$, can not be detected as a small sized
one, e.g. with $r_{hl}\sim 0.1''$.  The simulation shows that the
fraction of small size galaxies present in the HDF morphological
catalog is real and is not due to intrinsically larger objects which
have been shrunk by noise effects.

\section{THE DISTRIBUTION OF THE GALAXY SIZES IN LUMINOSITY AND REDSHIFT:
A COMPARISON WITH CDM MODELS}

We have computed the disk linear size $R_d$ and the absolute
blue magnitude for each galaxy in the catalog using the color
estimated redshifts as discussed in the previous section.

In Fig. 2 we plot the distribution of the observed sizes for the HDF
galaxies as a function of luminosity in four different redshift
intervals. The filled circles represent HDF spirals with
bulge-to-total ratio in the range $0.05<B/T<0.75$ while asterisks
represent galaxies with $B/T<0.05$, most of which of irregular
morphology (Marleau \& Simard 1998; Schade et al. 1998). HDF galaxies
with $B/T>0.75$ are excluded since they are bulge dominated
systems. The NTTDF galaxies of Paper I are also shown as empty
squares.

We also show in the shaded area the prediction of our rendition of the
standard semi-analytical CDM models. This relates the luminous
properties of galaxies to their circular velocity including the
hierarchical merging of dark matter halos, the merging of galaxies
inside the halos, the gas cooling, the star formation and the
Supernovae feedback associated with the galaxies.  Finally, the
circular velocity of the halos is connected to the disk scale length
using the Mo, Mao \& White (1998) model; the latter correlation
depends on the dimensionless angular momentum $\lambda$ whose
lognormal distribution $p(\lambda)$ is given in Mo et al. (1998). The
shaded area in Fig. 2 corresponds to that allowed for
$0.025<\lambda<0.1$, the values corresponding to the 10\% and 90\%
points of $p(\lambda)$.  The solid line corresponds to $\lambda=0.05$,
the 50\% point of $p(\lambda)$. The full $p(\lambda)$ distribution is
taken into account in the differential size distribution of galaxies
with $I<26$ (normalized to the total number) shown in Fig.~3 for
different redshift bins. A tilted CDM power spectrum of perturbations
with $n=0.8$ in an $\Omega=1$ Universe with $H_o=50$ km/s/Mpc has been
used.

The full details of our semi-analytic model are given in Appendix A of
Paper I together with the adopted set of star formation and feedback
parameters. The latter set was chosen as to optimize the matching
to the local I-band Tully-Fisher relation for bright galaxies and the
B-band galaxy luminosity function. Note that, since the disk velocity
is $\sim$ 20\% higher than that of the DM, a small offset ($\sim 0.5$
mag) between the predicted and the observed Tully-Fisher relation is
present (see Fig.~9 of Paper I).

Fig.~2 shows that at $z<1$ and for faint magnitudes ($M_B>-19$), the
observed sizes tend to occupy preferentially the small size region
below the 50\% locus of the angular momentum
distribution. Correspondingly, Fig. 3 shows the excess of small
($R_d<2$ kpc) size galaxies with respect to the CDM
predictions. Indeed, for $M_B>-19$, the predicted average disk size is
2.1 Kpc while the observed one is 1.4 Kpc. These results are similar
to those presented in Paper I, although extended down to $M_B<-16$ and
with a larger statistics.

The excess becomes less evident at brighter magnitudes, in agreement
with recent studies (Lilly et al. 1998, Simard et al. 1999) which
indicate little evolution in the morphological properties of bright
spirals in the CFRS sample up to $z\sim 1$. At $z\gtrsim 1$, the
larger statistics available with the present sample (with respect to
that used in Paper I) shows that the excess persists and involves
brighter galaxies ($M_B\gtrsim -20$) with an observed average
$R_d\simeq 1.3$ kpc respect to the predicted $R_d\simeq 1.9$ kpc. In
addition, the excess appears for all the galaxies in the sample
regardless of their morphological classification and so does not
depend on the selection procedure adopted for the spiral
sample. Furthermore, the above excesses cannot be due to the offset
(only $\sim 0.5$ mag) between the observed and the theoretical
Tully-Fisher relation, as confirmed by the good agreement of the
$R_d-M_B$ relation at the bright end.

In summary, Figs. 2 and 3 indicate that small-size galaxies appear
smaller and/or brighter than predicted by CDM at all $z$ (indeed, even
more at high $z$). Within the framework of the adopted standard
scenario of disk formation, which assumes the conservation of the
specific baryonic angular momentum (Mo, Mao \& White 1999), a viable
solution consists in introducing a brightening of small-size
galaxies. In particular, we note that shifting the predicted curves
toward higher luminosities results in a better fit to the data. At
$z\lesssim 1$ the best fitting shift is $\sim 1$ mag, while at larger
$z$ the best fitting shift is $\sim 1.5$ mag (Fig.~2).

\section{CONCLUSIONS AND DISCUSSION}

The present analysis performed on a larger and deeper sample, confirms
our previous findings at $z\lesssim 1$ (Paper I), where an excess of
faint ($M_B>-19$), small-size ($R_d<2$ kpc) galaxies with respect to
the CDM predictions was found.  The results presented here show that
the excess persits even at higher redshifts ($1<z<3.5$) and for
brighter galaxies ($M_B>-20$).  Several processes may be responsible
for the above excess (see Paper I), like the non conservation of the
gas angular momentum during the collapse in the Mo et al. model.
Alternatively, within the Mo et al. framework for disk formation, a
possible explanation can be sought in luminosity-dependent
effects related to the physical mechanisms involved in star formation
activity already at high $z$. Indeed, a shift of the shaded region
(the CDM prediction) by $\sim 1$ mag at $z< 1$ and 1.5 mag at $z>1.5$ is
sufficient to reconcile the CDM predictions with the observations.

Such shift could be due to the starbust brightening of the numerous
small size galaxies subject to close encounters/interactions.  This
brightening would have the advantage of explaining at the same time
the flat shape of the global cosmological star formation rate
$\dot{M}_*$ observed at $z>1.5$ in deep surveys (Steidel et al. 1999;
Fontana et al. 1999b) which results a factor $\sim 5-10$ higher than
predicted by CDM. Since the SFR is proportional to the UV-B
luminosity, a brightening of the predicted luminosities in small size
galaxies by $\sim 1-2$ mag is needed in both cases. The interaction
rate needed to reconcile the CDM evolutionary scenario with the
various observables can be derived by the following simple
considerations.  The SFR in a galaxy halo from a cool gas of mass
$M_{cool}$ can be written as $\dot{M}_*\approx
f_{gas}\,M_{cool}/\tau_i + (1-f_{gas})\,M_{cool}/\tau_*$, where
$f_{gas}$ is the fraction of gas converted in stars due to
interactions, $\tau_i$ is the timescale for interactions and $\tau_*$
is the quiescent star formation time scale. While only the latter term
is usually included in the CDM models, we note that for $f_{gas}\sim
0.1$ a $\tau_i$ shorter than $\tau_*$ by about a factor 100 would be
implied to obtain a SFR consistent with the high $z$ data; for the
population of small galaxies (dominating at high $z$) with a circular
velocity $v_c\lesssim$ 100 km/s (corresponding to $\tau_*\approx 5$
Gyr, see Cole et al. 1994) this would imply $\tau_i\approx 0.1$ Gyr,
in fact close to the dynamical time scale of these systems at $z\sim
2$. Note that for larger ($v_c>200$ km/s) disk galaxies, $\tau_*\sim
v_c^{-1.5}$ becomes smaller while $\tau_i\sim 1/N_g$ remains large due
to their small number density $N_g$ (Cavaliere \& Vittorini 1999), so
that the quiescent star formation mode prevails for these systems; in
addition since $N_g\sim (1+z)^3$ for all the galaxies, the interaction
time scale $\tau_i$ becomes ineffective at small $z$.

Detailed implementation of the interaction-driven star formation mode
in semi-analytical models will soon provide a more quantitative test
for the importance of this physical mechanism in determining the
galaxy properties at high $z$.

\newpage

\centerline {\bf FIGURE CAPTIONS}
\bigskip\bigskip
 
\noindent
Fig.~1. Deconvolved half light radii as a function of true values in
simulated data of the HDF. Error bars are one sigma confidence intervals.

\bigskip
\noindent
Fig.~2. Distribution of galaxies in the luminosity-size plane in four
redshift intervals. The disk radii are in kpc. Empty squares are NTTDF
galaxies; filled circles are HDF spiral galaxies with $0.05<B/T<0.75$;
stars are galaxies with $B/T<0.05$ most of which with irregular
morphology. The shaded region represents the region allowed by the
model. The upper and lower lines correspond to the 90\% and 10\%
points of the angular momentum distribution.  The solid line
corresponds to the 50\% point of the same distribution (see the text
for details).

\bigskip
\noindent
Fig.~3. Size distribution of the low and high luminosity spiral galaxies in
the HDF and NTTDF shown in Fig.~2. The corresponding curves are the
distributions predicted by the CDM model. An excess of small size
galaxies with respect to the CDM predictions is apparent at $R_d\simeq 1$
kpc.

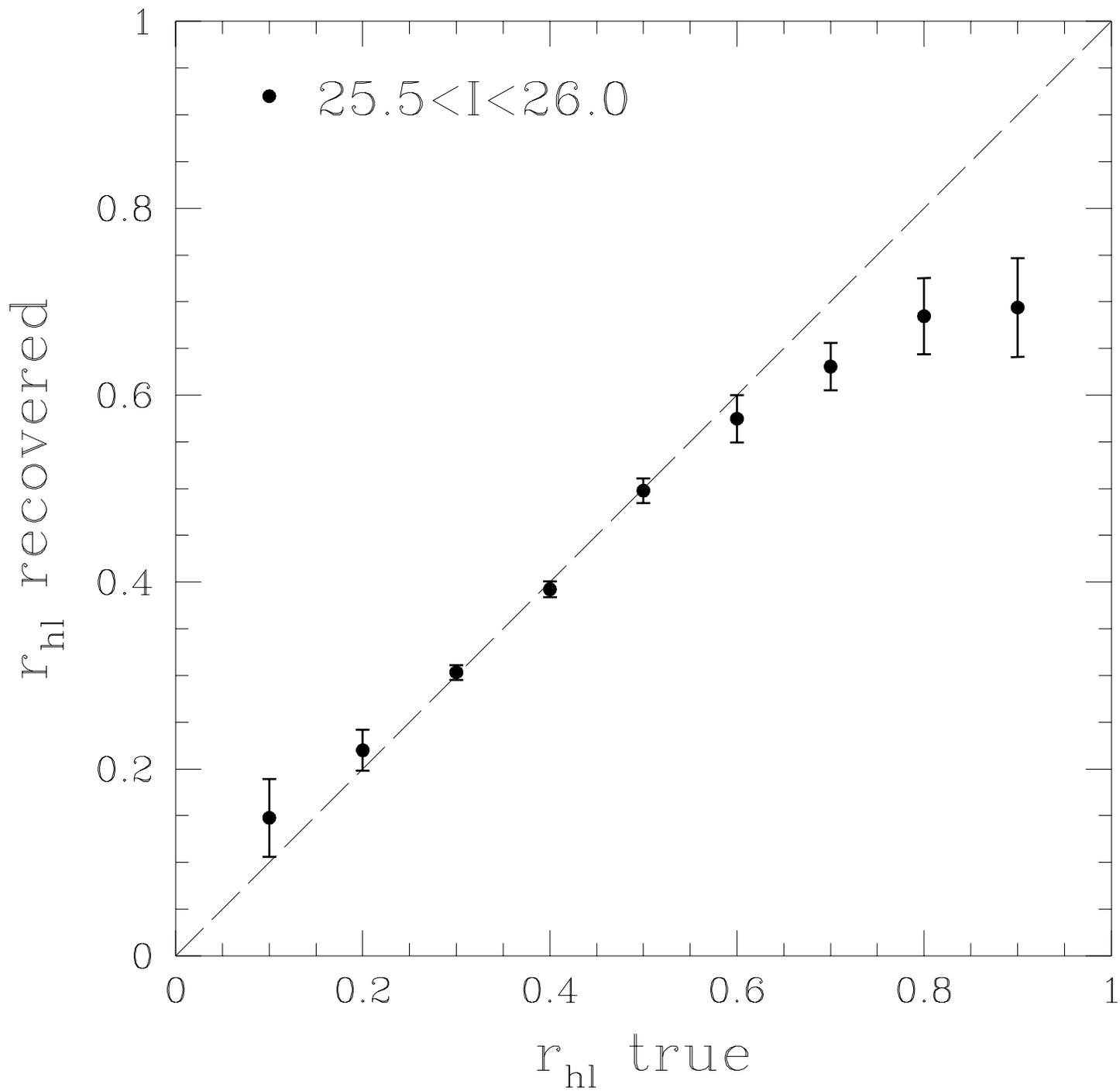
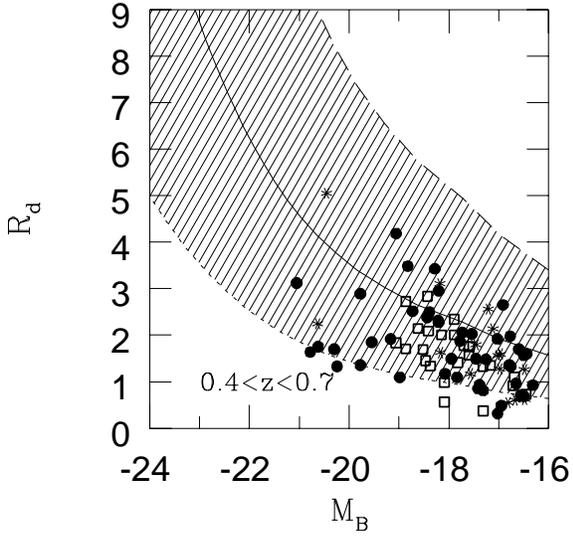
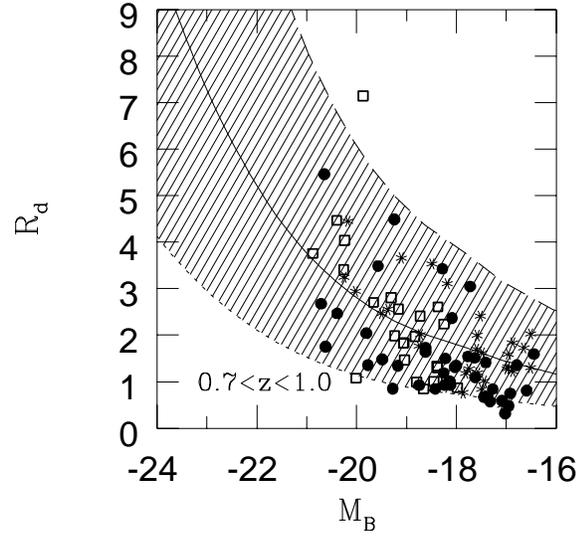
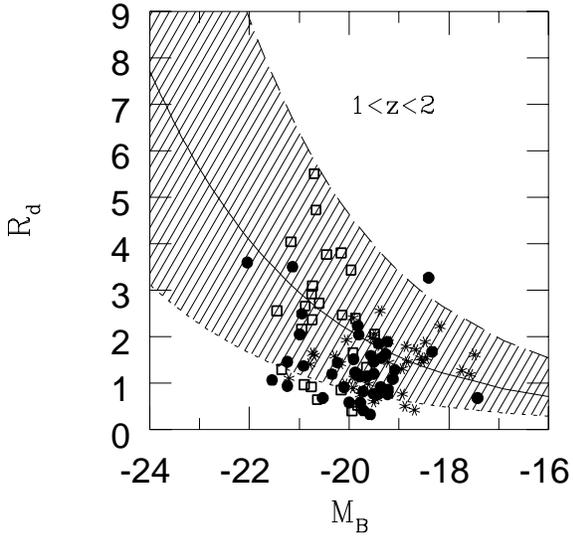
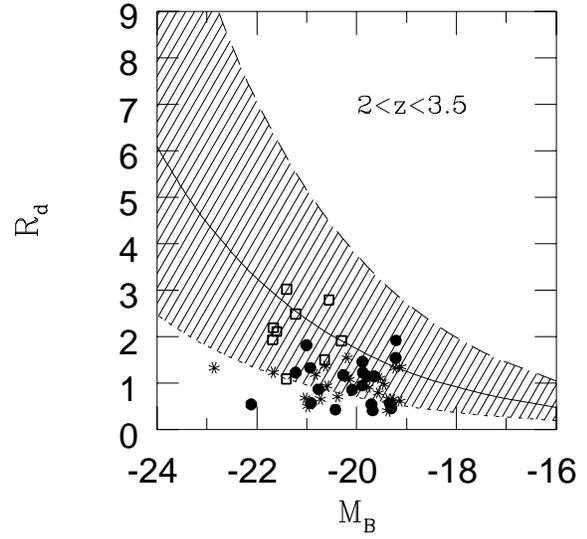
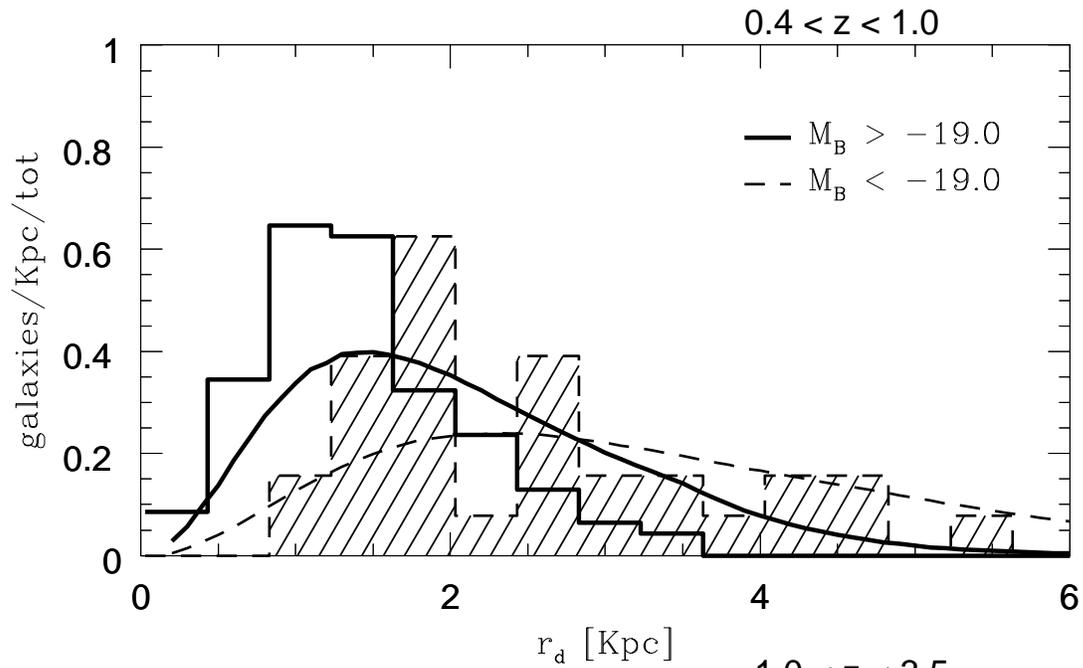
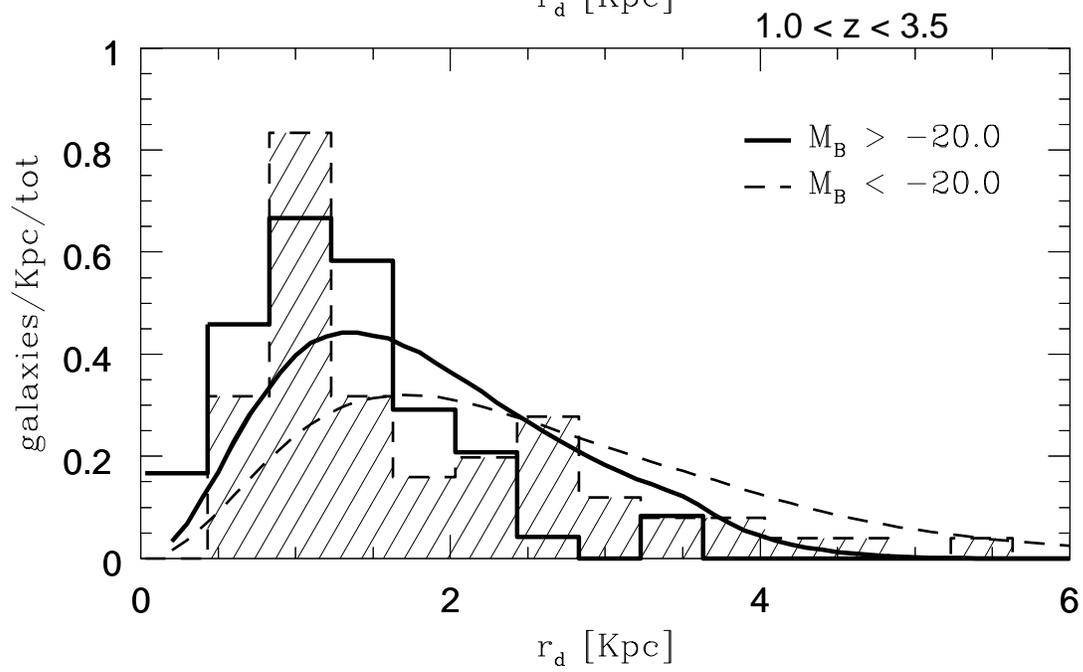
\end{document}